\documentclass[sigconf]{acmart}
\AtBeginDocument{%
  }

\usepackage{booktabs}
\usepackage{fontawesome}
\usepackage{soul}
\usepackage{float}

\newcommand{\insight}[1]{\noindent
\fbox{\begin{minipage}{0.465\textwidth}
\faLightbulbO\hspace{3pt}\textit{#1}
\end{minipage}}
}
\usepackage{balance}
\usepackage{enumitem}

\copyrightyear{2025}
\acmYear{2025}
\setcopyright{cc}
\setcctype{by}
\acmConference[]{}{}{}
\acmBooktitle{}

\renewcommand\footnotetextcopyrightpermission[1]{}

\begin{document}

\title[A Multi-Institutional Study of Student Views on Generative AI in Computing Education]{Excited, Skeptical, or Worried? A Multi-Institutional Study of Student Views on Generative AI in Computing Education}



\author{Isaac Alpizar-Chacon}
\authornote{Both authors contributed equally to this research.}
\email{i.alpizarchacon@uu.nl}
\orcid{0000-0002-6931-9787}
\affiliation{%
  \institution{Utrecht University}
  \city{Utrecht}
  \country{The Netherlands}
}

\author{Hieke Keuning}
\email{h.w.keuning@uu.nl}
\authornotemark[1]
\orcid{0000-0001-5778-7519}
\affiliation{%
  \institution{Utrecht University}
  \city{Utrecht}
  \country{The Netherlands}
}

\author{Imke de Jong}
\email{i.dejong1@uu.nl}
\orcid{0000-0003-0404-4011}
\affiliation{%
  \institution{Utrecht University}
  \city{Utrecht}
  \country{The Netherlands}
}

\author{Ioanna Lykourentzou}
\email{i.lykourentzou@uu.nl}
\orcid{0000-0002-4243-4128}
\affiliation{%
  \institution{Utrecht University}
  \city{Utrecht}
  \country{The Netherlands}
}

\author{Susan Rings}
\email{s.c.j.rings@uu.nl}
\orcid{0009-0002-3444-2301}
\affiliation{%
  \institution{Utrecht University}
  \city{Utrecht}
  \country{The Netherlands}
}


\begin{abstract}
The application of Artificial Intelligence, in particular Generative AI, has become more widespread among educational institutions. 
Opinions vary widely on whether integrating AI into classrooms is the way forward or if it is detrimental to the quality of education.
Increasingly, research studies are giving us more insight into the consequences of using AI tools in learning and teaching. Studies have shown how, when, and why students use AI tools. Because developments regarding the technology and its use are moving fast, we need frequent, ongoing, and more fine-grained investigation. One aspect that we do not know much about yet is how students use and think about AI across \textit{different types of education}. In this paper, we present the results of a multi-institutional survey with responses from 410 students enrolled in the computing programs of 23 educational institutions, representing high schools, colleges, and research universities. We found distinct usage patterns across the three educational institution types. Students from all types express excitement, optimism, and gratitude toward GenAI. Students in higher education more often report worry and skepticism, while high school students report greater trust and fewer negative feelings. Additionally, the AI hype has had a minimal influence, positive or negative, on high school students' decision to pursue computing. Our study contributes to a better understanding of inter-institutional differences in AI usage and perception and can help educators and students better prepare for future challenges related to AI in computing education.
\end{abstract}



\maketitle

\footnotetext{This is the authors’ version of the work. 
It has been accepted for publication in the \textit{Proceedings of the 25th Koli Calling International Conference on Computing Education Research, 2025}. 
The definitive version will be published by ACM.}

\vspace{2\baselineskip}

\section{Introduction}

The use of Generative Artificial Intelligence (GenAI) in computing education has been a major focus of both educational and computer science research, receiving increasing attention over the last three years. Research has explored how to embed GenAI tools into computing education \cite{prather2024beyond}, the capabilities of these tools for different types of programming tasks \cite{finnie2022robots,prather2023robots}, but also how students interact with and perceive these tools \cite{smith2024early,prather2023robots}. However, a question that remains open is how students use and think about GenAI across different types of education. Existing studies usually focus on a single type of education, often at the university level, overlooking other types of education, such as high schools and vocational colleges.

In this paper, we present the results of a large-scale, multi-institutional survey conducted across three types of educational institutions: high schools offering Informatics programs, colleges offering vocational programs focused on applied computing and Information and Communication Technologies (ICT), and research universities offering computing programs. The research question guiding our study is: \textit{How do students from different education types (high school, vocational college, research university) use and perceive Generative AI tools in computing education?}

A total of 410 students from 23 institutions in the Netherlands participated in our survey. Our study is exploratory in nature, aiming to uncover patterns in the usage and perception of GenAI among students across three distinct educational levels, given the known structural and pedagogical differences between these educational contexts. For example, while college and university students have experienced a longer period of education without exposure to GenAI tools, high school students have grown up with greater familiarity and accessibility to chatbots and other AI technologies. At the same time, high school students tend to be less set on specific professional paths or fields, where AI is increasingly prominent, and they are more at the beginning of exploring a broader span of future academic and career directions. In this light, it would be relevant to examine whether the emergence of GenAI is influencing their study choices. Differences also exist between vocational college students, who typically follow curricula with a stronger emphasis on the practical application of programming, and research university students, who often focus more on the theoretical foundations of computer science. These different educational objectives may, in turn, affect how these two student types engage with and perceive GenAI technologies.

This study makes two main contributions to the existing literature. First, it provides one of the first comprehensive comparative analyses of GenAI usage and perceptions across three distinct educational contexts, namely high schools, vocational colleges, and research universities. In doing so,  it addresses the relatively underexplored perspectives of high school and college students and situates the findings within the context of prior research.
Second, our study provides evidence-based insights from the student perspective. These insights can guide teachers and educational policymakers when designing education, and help those at college/university level understand the experiences and attitudes of students entering their programs from high school.

\section{Related work}
\label{related}

From the moment GenAI tools became easily accessible to the general public, the amount of research on this topic has been constantly increasing. In computing education research, studies have examined the use of GenAI tools in computing education (for overviews of different studies, see \cite{AGBO2025100266, 10883995, prather2023robots, prather2024beyond}).
Several dimensions of the topic have gained attention, such as the capabilities of Large Language Models (LLMs) in solving programming tasks (e.g.~\cite{finnie2022robots}), and the use of guardrailed LLMs in educational tools (e.g.~\cite{10.1145/3631802.3631830}).

One approach to studying the use of GenAI in programming is through experimental designs embedded in courses \cite{margulieux2024self,prather2024widening,app14104115,rahe2025how}. For instance, \citet{margulieux2024self} conducted a study with 40 students, collecting data at multiple points during an introductory programming course. Focusing on self-regulation and self-efficacy, the study found that students used GenAI to support learning rather than depend on it, with those showing higher confidence, lower fear of failure, or better grades using it less frequently and later in their learning process. Most recently, \citet{rahe2025how} conducted a study with 37 programming students using a ChatGPT-powered chatbot that logged all their interactions while working on code-writing tasks. They found that the chatbot was not used universally and was primarily employed for learning and code generation.

Other approaches to understanding the use and perceptions of GenAI involve collecting self-reported data. There have been several large- and small-scale studies of student perceptions regarding GenAI, including studies aimed at computing programs. Shortly after the introduction of ChatGPT, \citet{smith2024early} and \citet{prather2023robots} examined university students' perceptions and use of these tools. \citet{smith2024early} found that students at a university in the United States were at that time using GenAI tools for learning, writing, and coding, but none of the students indicated they used GenAI to generate complete solutions to assignments. The study also showed that students who used the tools more often were more positive about the benefits of GenAI and were more likely to say the use of the tools should be encouraged or conditionally encouraged. \citet{prather2023robots} conducted an international survey that confirmed that students use GenAI for generating text and coding. They also inquired whether students feel GenAI tools can guide students as effectively as a teacher can. Students indicated this is not the case. Finally, 95\% of the student respondents did not find it ethical to auto-generate complete assignment solutions.

\citet{keuning2024students} took a longitudinal approach and surveyed the use and perception of GenAI in programming courses at a Dutch university through three consecutive surveys in the academic year 2023--2024.
The results showed that the use of GenAI tools had increased overall. GenAI was, for example, gaining preference as a help-seeking strategy, with MSc students ranking it higher than BSc students. The increasing preference to use GenAI as a help-seeking resource was also noted by \citet{hou2025evolving}, who surveyed North American university students during roughly the same time period. In the study of \citet{keuning2024students}, some concerns about the ethical use of the tools arose from the results. Over the course of the study, there was an increase in the number of students who considered it ethical to auto-generate a complete solution for an assignment and submit it without understanding it. For courses that use programming as a tool, students placed more emphasis on GenAI being part of the future, therefore justifying its use. Students also showed a critical view on the tools, however. For core programming courses, they emphasized the importance of learning and the need to disallow GenAI as they believed it hinders their learning process. 
In an interview study by Zastudil et al. \cite{zastudil2023generative}, this critical view was echoed by 12 computer science students from a research university in the United States. The interviewed students remark that students might become over-reliant on the tools or that they provide incorrect or misleading information. These students also note that the tools can be very useful when used correctly, for example, to reduce effort spent on tasks seen as ``busy work'' or to generate additional examples and explanations.

A recent article discusses the role of GenAI in social learning and help-seeking \cite{hou2025roads}. In this exploratory study, North American undergraduate computing students were interviewed on this topic, uncovering major concerns such as reduced help-seeking with peers, problems with motivation, and students feeling isolated. The authors conclude by stating, despite the study's limitations that should be addressed in follow-up studies, that ``AI may be inadvertently destroying the very social fabric that supports meaningful learning.''
In an earlier study, some of the same authors also surveyed and interviewed students on help-seeking behavior in the same North American context \cite{hou2024effects}. They found that students prefer different types of help resources for different tasks. For example, after online search, friends and teachers were most important for learning about concepts, followed by GenAI. They also discuss factors such as trust, quality, and timeliness.

Fewer studies are looking at the use and perceptions of students in high school settings. \citet{LEE2024100253} surveyed high school students in the United States in 2023 to discover how often they use tools like ChatGPT, and whether they believe using these tools should be permitted (and for which tasks). The results showed that these tools were mainly used by students to explain concepts and generate ideas. Completing (parts of) assignments with a tool was done less often, although students in public schools reported doing this more often than students in private schools. The tools were also used for programming, but only to a limited extent. When asked for what tasks the use of AI should be allowed, using it to generate ideas and explain concepts was mostly supported. Most students also agree that the tools should not be used to generate complete solutions to assignments. For programming, more than half of the students indicated it should always or sometimes be allowed.

\section{Method}
\label{method}

\subsection{Survey design}
\label{sect:surveydesign}
Our survey was adapted from the 2023 ITiCSE Working Group survey on Generative AI~\cite{prather2023robots}, which has been extended in subsequent studies \cite{keuning2024students}. 
We added demographic questions, removed a few questions that were irrelevant for our context, and incorporated questions from a large-scale recent survey conducted in Australia~\cite{Chung2024}, as well as questions on perceived learning from another study ~\cite{alpizar2025students}. The final instrument comprises two parts: Part 1 (general) includes questions on demographics and students' perceptions of GenAI, and Part 2 (course-specific) includes questions on students' use of GenAI in a specific course, where answers are expected to vary for the different courses. 
Most questions are multiple choice, but we also added some open questions to get qualitative student insights on (1) whether GenAI should be permitted in courses/exams and why, (2) if and how students’ views on GenAI have changed over time, and (3) if and how students perceive that their learning process was affected by the use of GenAI.
Because our target population includes students from high schools, an education expert reviewed the survey for suitability, and a former high school teacher checked the phrasing and language. We believe that the resulting changes to questions improved clarity for all participants.

The final survey is provided online.\footnote{\url{https://doi.org/10.5281/zenodo.17144157}}
It was created in Qualtrics and made available in two languages: Dutch and English. The survey took between 10 and 20 minutes to complete. Participants could enter a raffle to win one of several 20 euro gift vouchers as participant compensation. 
Their email addresses were stored separately from their responses, ensuring that their answers remain anonymous and could not be linked to them. 

\subsection{Participants}

The study design was approved by the Ethics Review Board of our university. 
Participants comprised three student groups studying computing:

\begin{itemize}
    \item \textbf{High school students}. In the Netherlands, some high schools offer an elective Informatics program for the higher grades. We target these schools and students. We require participating students to be 16 years or older, the age at which they can decide for themselves to participate. 
    \item \textbf{College students}. We target students from colleges enrolled in computer science-related programs. In the Netherlands, these institutions focus on vocational education, preparing students for industry roles in ICT, and offer mostly Bachelor's programs, and some applied Master's programs.
    \item \textbf{University students}. These students are enrolled in a computing program at a research-oriented university that typically offers Bachelor's and Master's degrees. Non-native students also often participate in these programs; therefore, we included the option to complete the survey in English.
\end{itemize}

\subsection{Distribution}
Given our aim to collect responses from a wide range of institutions, we invited teachers and education coordinators from other institutions to help with the distribution of our study to their students. We reached those individuals through three main channels: 

\begin{itemize}
    \item Personal contacts of the research team.
    \item National mailing lists. Our request was announced on a high school informatics mailing list, a mailing list for vocational study programs in ICT, and a mailing list for computing education researchers in the Netherlands.
    \item The research team posted the request on Social Media.
\end{itemize}

Our request included an information letter about the study, and 
email template invitations in two languages that they could share with their students.
We also offered institutions the option to receive an individualized report of their students' responses, provided the number of these responses was sufficient to ensure anonymity.

\subsection{Data processing}
\subsubsection{Collection and privacy}
At the beginning of the survey, participants could read an information sheet outlining the study’s purpose, research team, and contact details for questions or complaints. The sheet also specified that no identifiable or sensitive personal data would be collected, the results would be used solely for dissemination and scientific purposes, and that the data would be stored in secure servers for a minimum of 10 years.

\subsubsection{Cleaning}
We kept all responses that completed at least the first part of the survey. We removed two pairs of nearly identical responses that were submitted almost simultaneously, as they were probably duplicates.  
Finally, we checked response consistency and corrected the institution type for two participants whose selection did not match the institution name.

\begin{table*}[!t]
\caption{Participating institutions and programs.}
\label{tab:inst}
\begin{tabular}{lrrrrrrrrrrrrrrrrrr}
\toprule
\textbf{Type} & 
  \textbf{Institutions} & \textbf{Students}
    & \multicolumn{3}{l}{\textbf{Grade}}
    & \multicolumn{6}{l}{\textbf{Bachelor's}}
    & \multicolumn{6}{l}{\textbf{Master's}} \\
    &&&
    Y4 & Y5 & Y6 &
    Y1 & Y2& Y3 & Y4 & Y5+ & \textit{Tot.} &
    Y1 & Y2& Y3 & Y4 & Y5+ & \textit{Tot.} \\ \midrule

High school & 7 & 86 (21\%) & 39 & 35 & 12 \\
College & 7 & 203 (49.5\%) &&&& 50 & 67 & 33 & 29 & 15 & \textit{194} & 4 & 2 & 1 & 0 & 2 & \textit{9} \\
University & 9 & 121 (29.5\%) &&&& 11 & 43 & 30 & 7 & 5 & \textit{96} & 15 & 5 & 2 & 1 & 2 & \textit{25} \\ \bottomrule
\end{tabular}
\end{table*}

\subsubsection{Analysis}
We analyzed the closed questions using a comparative analysis that examined the differences and patterns among the three educational groups.
Given that most of our data was ordinal (Likert-scale), 
we used non-parametric tests, namely the Kruskal--Wallis H test to detect overall group differences and Mann--Whitney U tests for pairwise comparisons. 
To examine associations between categorical responses, we employed the Chi-Square test. To explore the 
relationships between GenAI usage and reported perceptions and attitudes, we conducted Spearman correlation analyses. Given the exploratory nature of our analyses and the many comparisons performed, we consistently applied the Bonferroni correction.

We analyzed the three open questions using deductive and inductive coding.
First, we categorized the responses deductively (e.g., using the ``yes'', ``no'' and ``under certain conditions'' categories for the question on GenAI permissibility).
Then, we applied inductive coding to analyze student rationales. For the first two open questions (on the permissibility of GenAI in courses/exams and on the change of student views on GenAI over time), two researchers independently performed the deductive coding, and then resolved differences in coding through discussion. The third question was coded by one researcher.

\section{Results}
\label{results}

Our final dataset includes 410 responses to the survey. Most participants (87\%) completed both parts of the survey, resulting in 356 full responses and 54 responses to only the first part. 

\subsection{Demographics}
Participants come from 23 different institutions: 7 high schools, 7 colleges, and 9 research universities (Table~\ref{tab:inst}). 
College students account for almost half (49.5\%) of the dataset, followed by research university students (29.5\%), and high school students (21\%).

The majority of college and university students are enrolled in Bachelor's programs. 
The high school students are from the final three years of high school (Y4 -- Y6), all following the Informatics program for two types of high school education: ``senior general secondary education'' (5-year program, preparing for college, 25 respondents), and ``pre-university education'' (6-year program, preparing for university education, 61 respondents).
Among the college and university respondents, there are 281 domestic students and 43 international students. Students come from a wide range of computing programs, such as Computer Science, Software Engineering, Data Science, Information Sciences, Bio-informatics, Creative Media and Game Technology, Embedded Systems Engineering, Artificial Intelligence, and ICT business. A small number of students are from non-computing majors and are following a computing course.

Participants comprise 74 females, 319 males, 4 non-binary individuals, 6 participants who self-described, and 7 who did not indicate their gender.
Of the students in higher education, 85 (26\%) are the first person in their family to study at a college or university, 227 (70\%) are not the first, and 12 (4\%) prefer not to disclose.

\subsubsection*{Programming proficiency and experience}

High school students estimated their programming proficiency to be 2.4 on average (1--5 scale), and college and university students both estimated it at 3.3.
We also asked them to estimate how important they think programming will be for their future career  (1--5 scale). College students rated this the highest with 4.2 on average, followed by university students with 3.8, and high school students with 2.5.
We asked if students have done programming outside of their educational institution. As expected, higher education students  reported programming as a hobby more often (college 65\%,  university 60\%), and/or as a job (college 25\%, university 29\%). About one-third of these students do not program outside their studies.
Among high school students, this proportion is over two-thirds, with fewer doing programming as a hobby (28\%) or as a job (2\%).

\subsection{Usage}
\insight{\textbf{Key insights:}
\begin{itemize}[leftmargin=*,labelsep=5pt]
    \item High school students use GenAI mainly for text tasks; in higher education, usage shifts toward programming, especially among college students. 
    Learning is a common use across all groups.
    \item As their use of GenAI increases, higher education students show less concern about its impact on learning and greater confidence in its guidance.
\end{itemize}
}

\begin{figure}[!bp]
    \centering
    \includegraphics[width=0.47\textwidth]{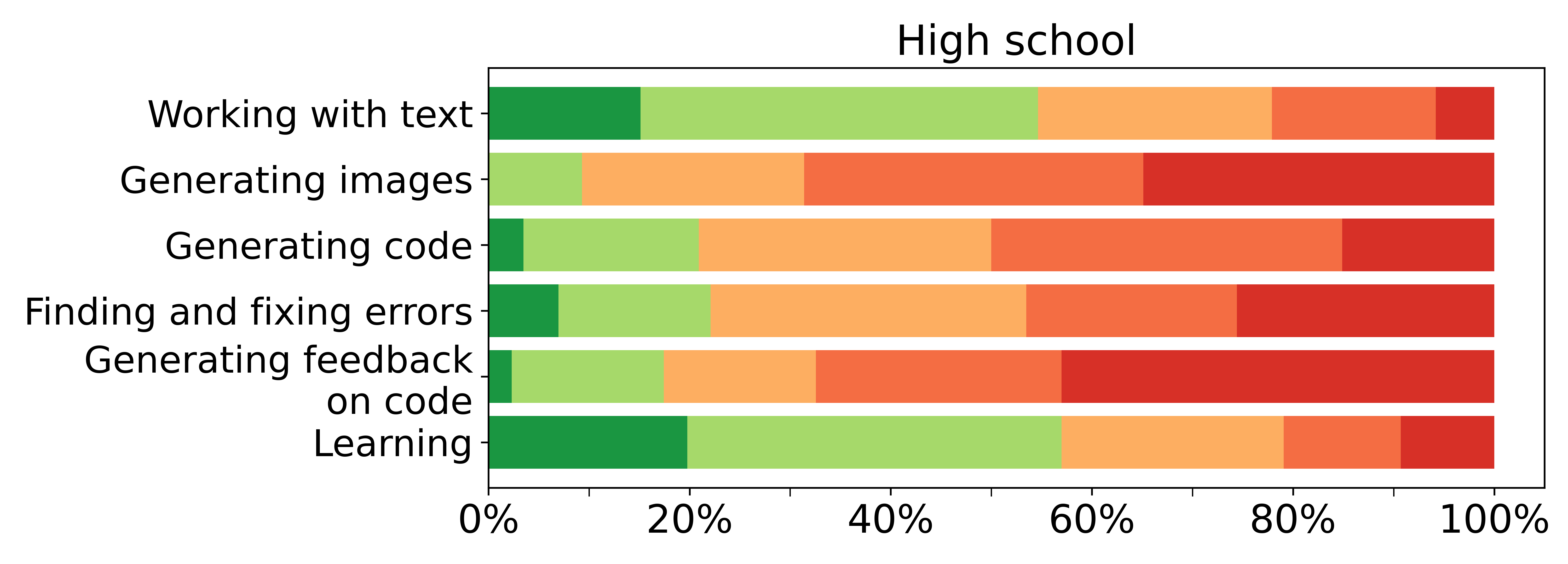}\\[0.2em]
    \includegraphics[width=0.47\textwidth]{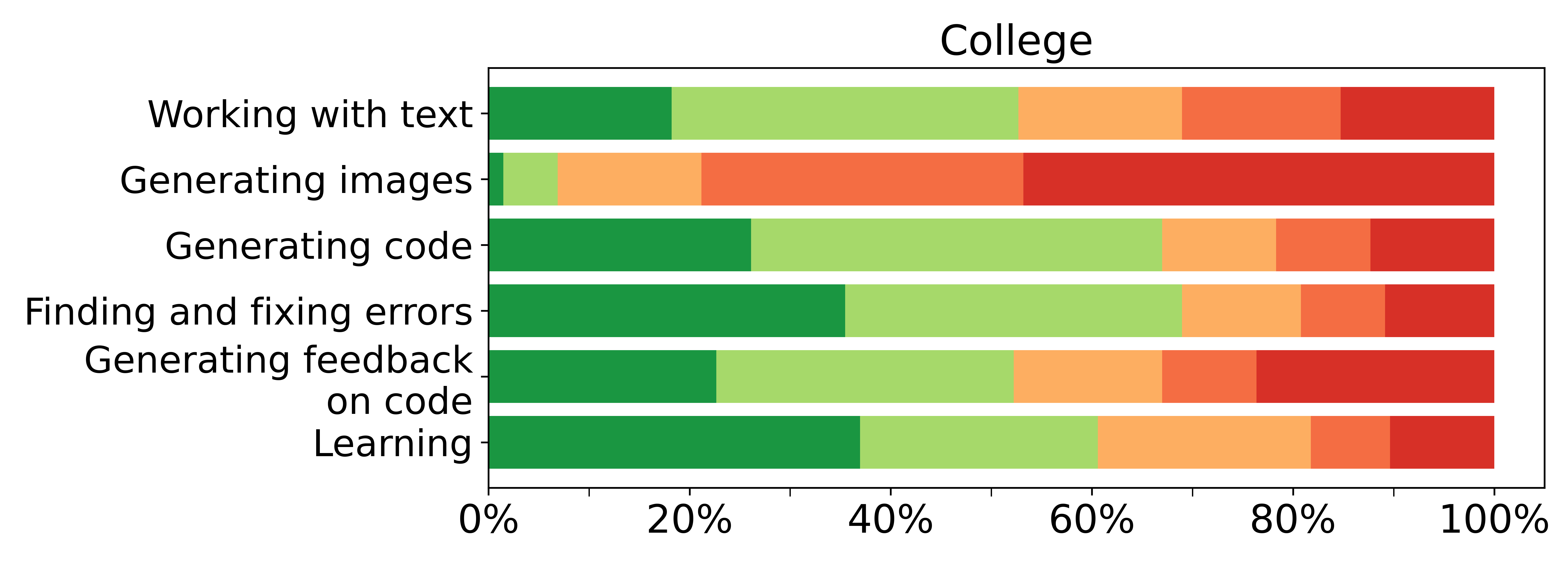}\\[0.5em]
    \includegraphics[width=0.47\textwidth]{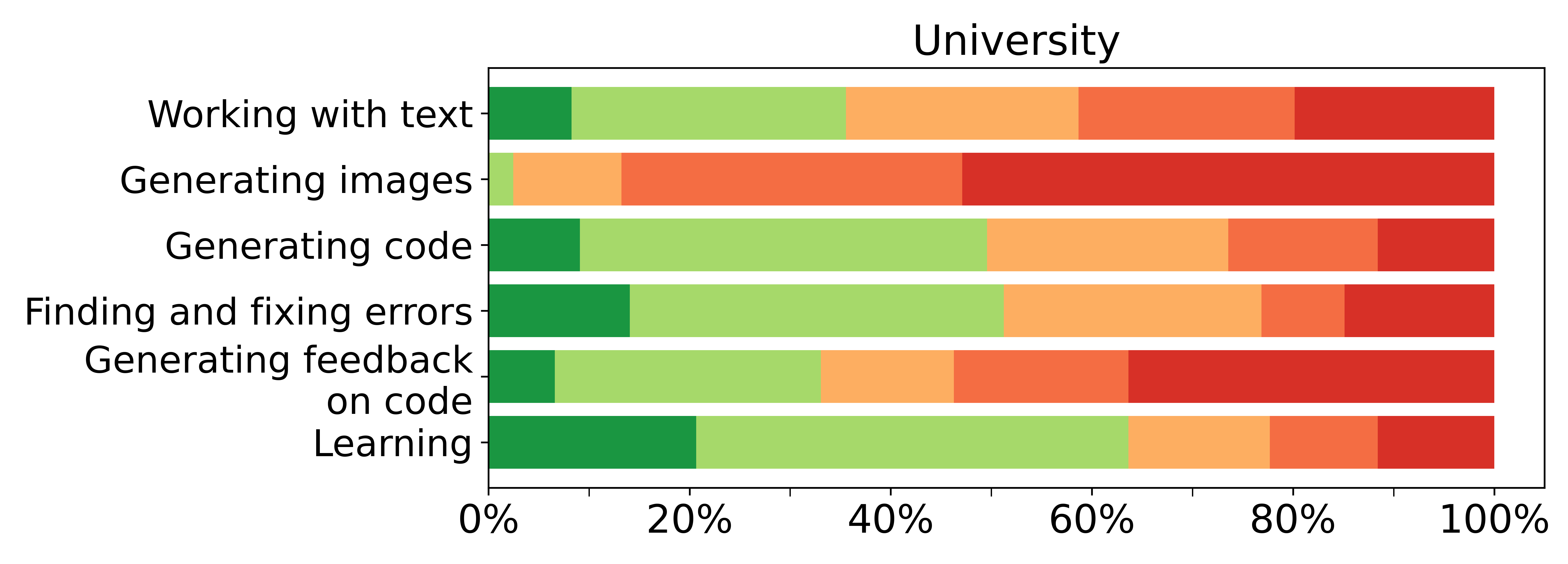}\\[0.5em]
    \includegraphics[width=0.4\textwidth]{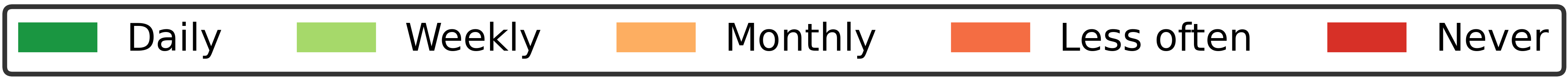}
    \caption{Reported GenAI use frequency across tasks.}
    \label{fig:usage_3}
\end{figure}

\begin{figure} [!tbp]
    \centering
    \includegraphics[width=1\linewidth]{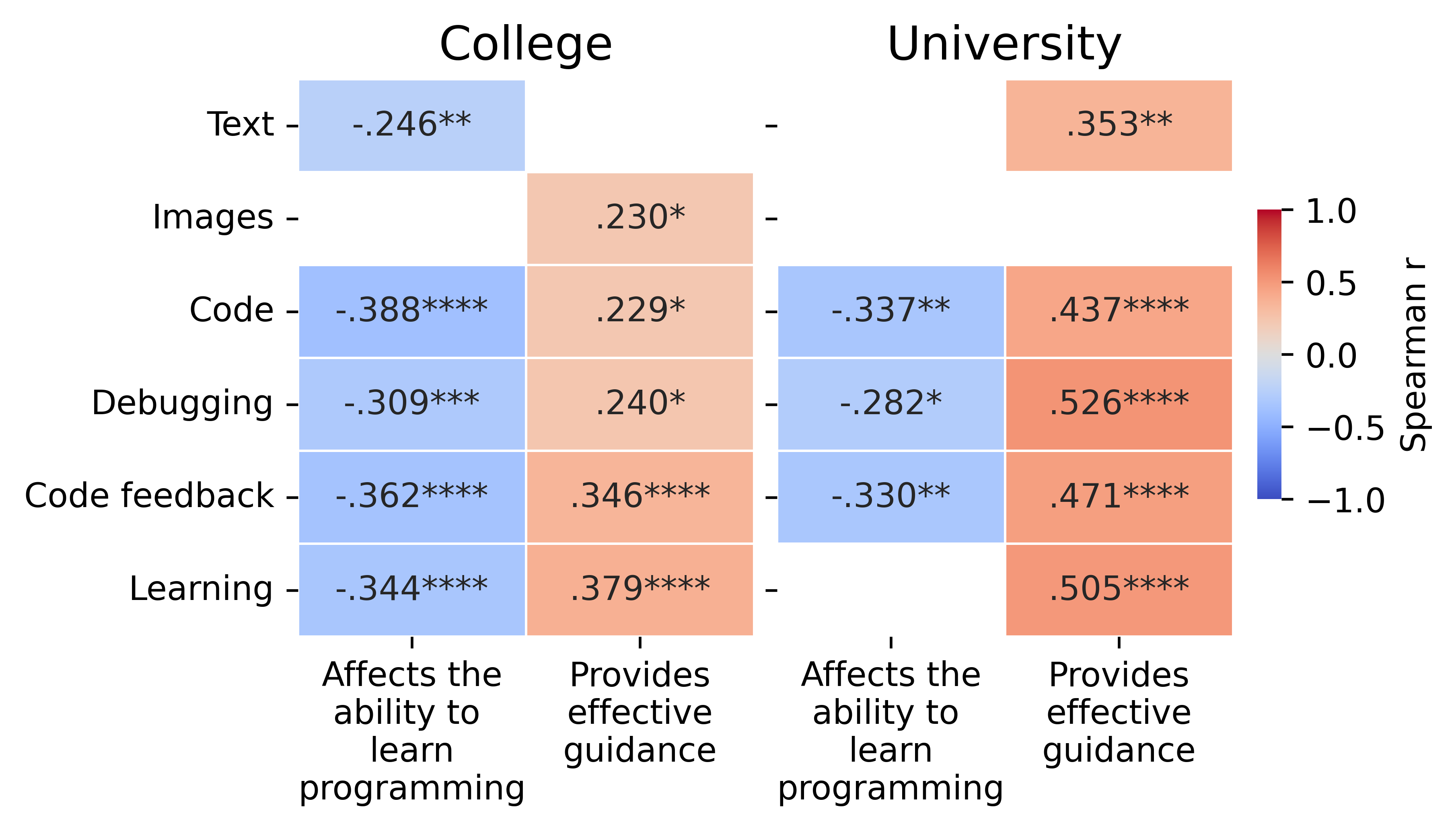}
    \caption{
        Significant Spearman correlations between GenAI use and implications, reported separately for college and university students. \\
        \textit{Note.} $^*$ $p < .05$, $^{**}$ $p < .01$, $^{***}$ $p < .001$, $^{****}$ $p < .0001$
        }
    \label{fig:usage_implications_correlations}
\end{figure}

Generally, the three educational groups show distinct usage patterns, as seen in Figure \ref{fig:usage_3}. High school students reported the most frequent use when working with text (54.6\% at least weekly) and learning about topics and concepts (57\% at least weekly). In contrast, they reported infrequent usage for programming-related tasks—for instance, 79.1\% generate code no more than once a month. This result is expected, as students typically receive only a few hours of Informatics instruction per week. College students reported the most frequent use of GenAI for programming-related tasks among all three groups. For example, 69\% reported at least weekly GenAI usage for finding and fixing bugs. This elevated usage was confirmed by multiple pairwise comparisons using Mann--Whitney U tests, with significant differences observed between college and the other groups across all three programming-related tasks (e.g., generating code: college vs. university, \(U = 14899.5\), \(p_{adj} = .002\)). University students reported frequent use of GenAI—at least weekly—for generating code (49.6\%) and identifying or fixing errors (51.2\%). All three groups reported a similar, very frequent use of GenAI for learning, with an average of 60.4\% using it at least weekly. A Kruskal--Wallis H test showed no statistically significant difference among the groups, \( H(2) = 4.16 \), \( p = .125 \).

We asked students to rate three statements about the implications of using GenAI. High school students were more likely to disagree (51.2\% somewhat or strongly) and less likely to agree (20.9\% somewhat or strongly)
that frequent GenAI use for code generation hinders learning, compared to college (38.4\% disagree, 43.4\% agree) and university students (38.8\% disagree, 49.6\% agree). Mann--Whitney U tests confirmed these differences (high school vs. college, \(U = 7060\), \(p_{adj} = .025\); high school vs. university, \(U = 3936\), \(p_{adj} = .002\)). 
A similar pattern appears for the statement about GenAI providing guidance as effectively as human teachers: high school students were more likely to agree (40.7\%) and less likely to disagree (31.4\%) than college (32.5\% agree, 47.8\% disagree) and university students (37.2\% agree, 42.2\% disagree), with a significant difference between high school and university students (\(U = 10354\), \(p_{adj} = .031\)).
Finally, regarding concerns about overreliance on GenAI, high school students showed the least agreement (27.9\%), followed by university (39.6\%) and college (41.9\%) students. However, a Kruskal--Wallis H test found no significant difference among the groups (\(H(2) = 5.19\), \(p = .075\)).

We further explored how the use of GenAI for the different tasks relates to students’ views on its potential implications. A Spearman correlation analysis revealed a moderate positive association between the use of GenAI when working with text and the worry of overreliance in high school students, \(r_s = .328 \), \(p = .037\). This suggests that students who use GenAI more for generating text worry about becoming over-reliant on these tools. For higher education students, we observed a weak to moderate negative association between all types of GenAI usage for programming and students' agreement that it hinders their ability to learn programming. This suggests that more frequent usage is linked to less concern about its impact on learning. Additionally, there is a weak to strong positive association between most types of GenAI usage and students' agreement that GenAI can provide effective guidance, indicating that more frequent use is associated with greater confidence in its usefulness. The correlation values are shown in Figure~\ref{fig:usage_implications_correlations}.

\subsection{Policies}
\insight{\textbf{Key insights:}
\begin{itemize}[leftmargin=*,labelsep=5pt]
    \item College students report more permissive institutional GenAI policies compared to high school and university students. 
    \item Across all educational types, most students believe GenAI should be allowed to support learning, but not for exams or graded tasks.
\end{itemize}
}

When asked if the GenAI policies at their institutions are clear, high school students were more likely to somewhat or strongly disagree (36.1\%) and less likely to somewhat or strongly agree (40.7\%), compared to students in college (20.2\% disagree, 58.6\% agree) and university (23.1\% disagree, 52.9\% agree). Mann--Whitney U tests confirmed these differences (high school vs. college, \(U = 6557\), \(p_{adj} = .001\); high school vs. university, \(U = 4123\), \(p_{adj} = .024\)).

\begin{figure}[bp]
    \centering
    \includegraphics[width=1\linewidth]{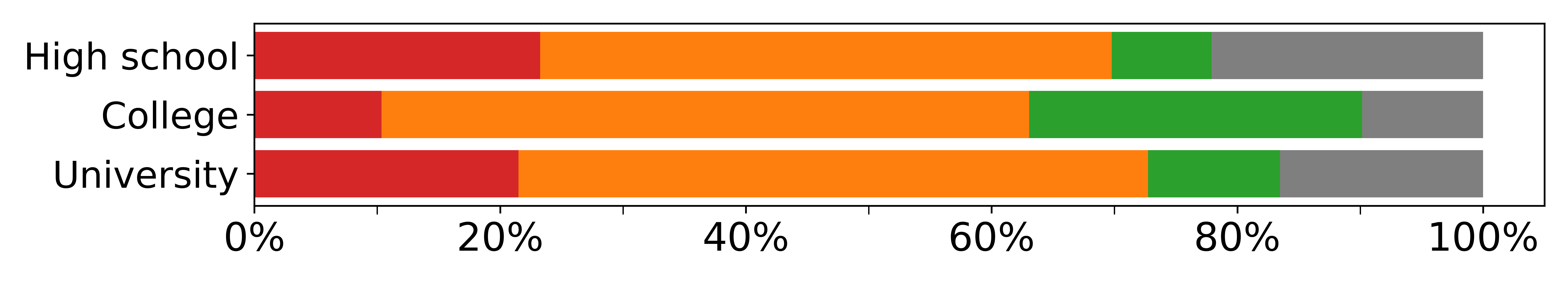}\\[0.5em]
    \includegraphics[width=0.4\textwidth]{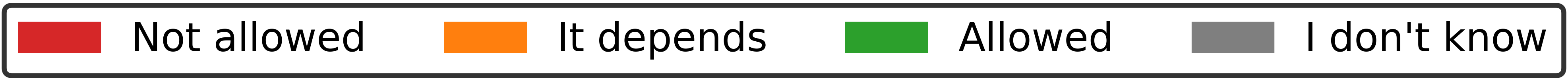}
    \caption{Allowance spectrum of GenAI use.}
    \label{fig:policies}
\end{figure}

Students also reported on the general policy regarding the use of GenAI at their institutions, as shown in Figure~\ref{fig:policies}. High school and university students reported a similar allowance spectrum. In contrast, college students reported fewer instances of disallowance (10.3\%), more cases of allowance (27.1\%), and were less likely to indicate that the policy was unknown (9.9\%). Pairwise Chi-Square tests confirmed these differences between high school and college students, \(\chi^2(3) = 24.38\), \(p_{adj} < .001\), and between college and university students, \(\chi^2(3) = 18.91\), \(p_{adj} < .001\). No significant difference was found between high school and university students.

Next to their perception of what is or is not allowed, we also asked whether students themselves feel the use of GenAI should or should not be allowed. We distinguish between students who say they believe GenAI should be allowed, those who say it should not be allowed, and those who state it should be allowed given specific conditions. The pattern in the 384 responses is the same for all educational types. The majority of students indicated that the tools can be used under certain conditions, providing different reasons. Students feel that they should be able to use the tools to support their learning process, but not during exams or for assignments that are graded. A college student said: \textit{``I would allow it for courses (in the classroom) because it can help with understanding problems and providing practice exercises. I would not allow it for exams because this is about the student's knowledge and skills, not the AI.''} Students also mentioned that it should be dependent on the course or assignment goals. As a university student noted: \textit{``[...] if the students are already able to program, I would allow the use of GenAI to generate small pieces of code that are used in the bigger code. For students that are learning programming, I think using GenAI for everything will not help them learning programming. ''}

For each educational group, the second most common reaction was that using GenAI should \textit{not} be allowed. These students claimed that using these tools does not support the learning process, but rather hinders the development of knowledge and skills. A university student said: \textit{``No. GenAI is a useful tool, but it takes away the most challenging part of learning new programming skills - debugging and becoming self sufficient. Uni courses/exams are designed in a way to guide you through the problem solving and learning goals; using LLMs skips most of these lessons and directs you linearly to the answer. This makes students reliant on the problem-solving abilities of LLMs, instead of developing these skills themselves.''} Other replies indicated it hinders the testing of this knowledge or these skills when the tools are used for exams or assignments. Students also remarked that the tools cannot be incorporated into the current educational setup or that there are societal problems related to their use, such as the level of energy consumption and copyright infringements. 
Finally, for all types of education, there are some students who indicate that it should always be allowed. Most cited reasons for this are that these tools are used in the careers they are preparing for and that this is the technology of the future, which they should learn to use. As a high school student stated: \textit{``Yes, using AI is probably a big part of the future so it's better to make sure people know how to use GenAI properly.''} Students also remarked that the tools can provide additional explanations when a topic is unclear.  

\subsection{Ethical aspects}
\insight{\textbf{Key insights:}
\begin{itemize}[leftmargin=*,labelsep=5pt]
    \item High school students report higher perceived misuse of GenAI by themselves and their peers than higher education students. 
    \item High school and college students hold more permissive ethical views on using GenAI for assignments than university students.
\end{itemize}
}

\begin{figure}[!tbp]
    \centering
    \includegraphics[width=0.47\textwidth]{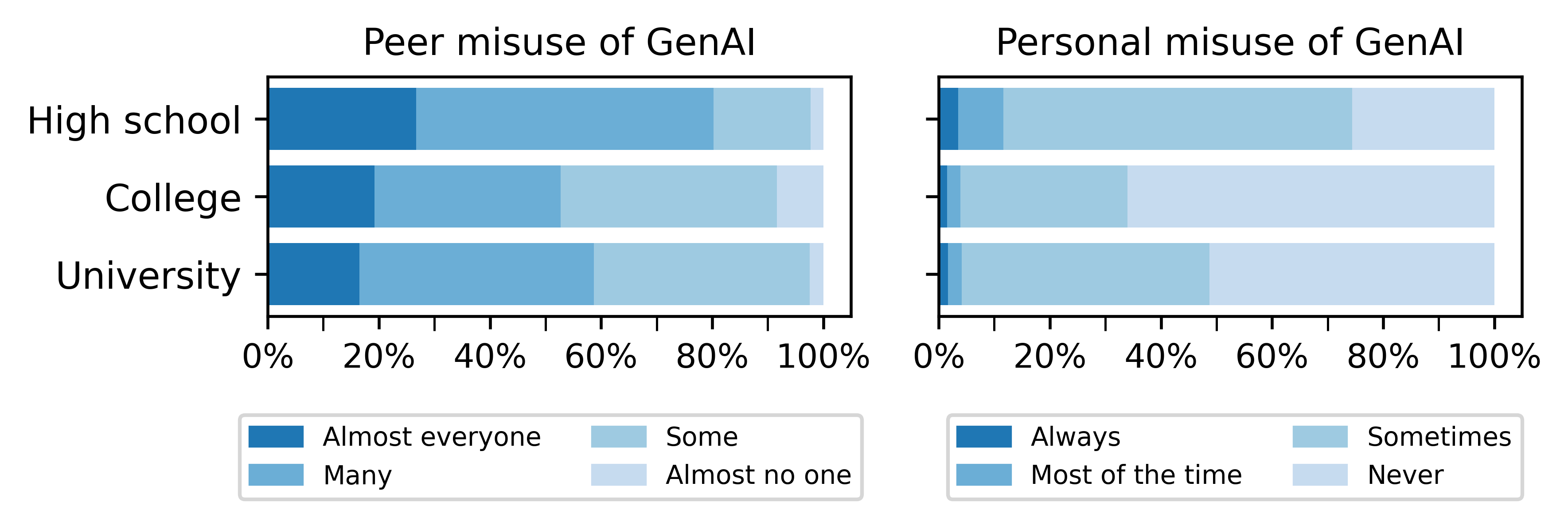}\\[0.5em]
    \includegraphics[width=0.47\textwidth]{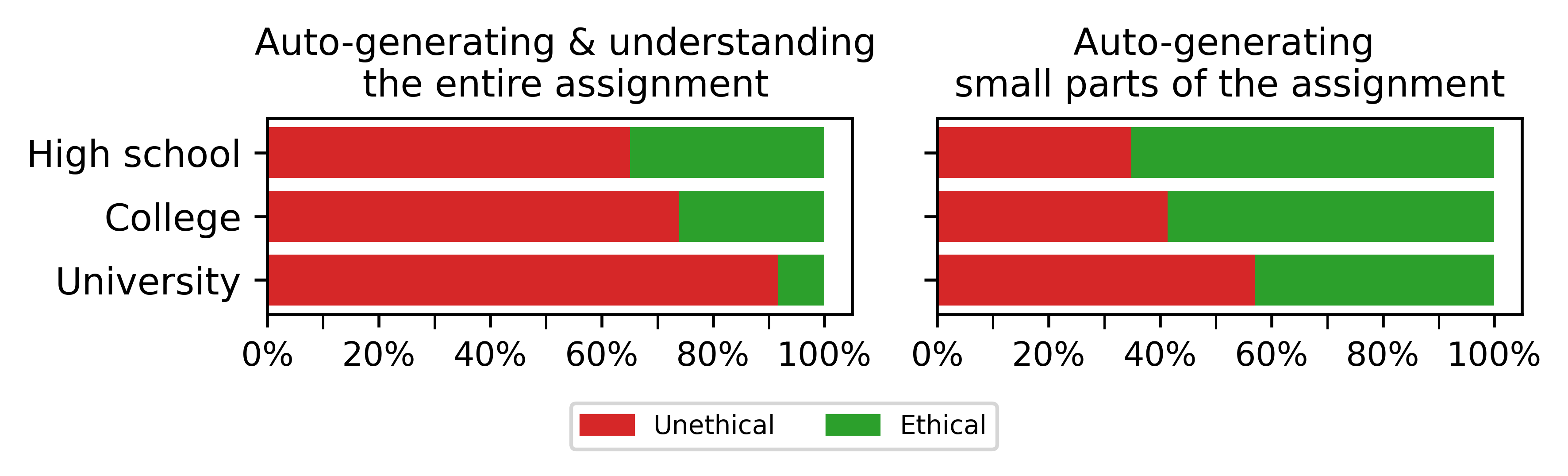}
    \caption{GenAI misuse (top) and perceived ethics across two use cases (bottom).}
    \label{fig:ethical_comb}
\end{figure}

We asked students several questions related to the ethical use of GenAI. As shown in Figure~\ref{fig:ethical_comb} (top left), high school students reported a higher perceived misuse of GenAI among their peers compared to college and university students. In high schools, students believe that either almost everyone (26.7\%) or many (53.5\%) of their classmates use GenAI in ways their instructors would not approve. The combined share of these two categories is notably lower among college (52.7\%) and university (58.6\%) students. Mann--Whitney U tests confirmed the higher belief of unethical use in high school (high school vs. college, \(U = 11065\), \(p_{adj} = <.001\); high school vs. university, \(U = 6426.5\), \(p_{adj} = .0.006\)). A similar trend appears when students were asked whether they had personally used GenAI when they were not supposed to (Figure~\ref{fig:ethical_comb}, top right). High school students reported the highest level of misuse, with only 25\% indicating they have never done so—a stark contrast to 66\% of college students and 51.2\% of university students who report never using GenAI inappropriately. Mann--Whitney U tests again showed significant group differences (high school vs. college, \(U = 12344\), \(p_{adj} = <.001\); high school vs. university, \(U = 6670.5\), \(p_{adj} = <.001\)).

In another question, we asked students to evaluate whether certain GenAI-related actions were ethical or not. For five out of the seven situations, pairwise Chi-Square tests showed similar response distributions across all educational groups. However, in two cases, high school and college students showed a more permissive stance, labeling actions as ethical more often than university students, as shown in Figure~\ref{fig:ethical_comb} (bottom). The number of students who considered it ethical to auto-generate an entire assignment and submit it after understanding it is 34.9\% and 26.1\% for high school and college students, respectively, while that number decreases to 8.3\% in the university (high school vs. university, \(\chi^2(1) = 21.17\), \(p_{adj} = <.001\); 
college vs. university, \(\chi^2(1) = 14.29\), \(p_{adj} = <.001\)). A similar pattern appears for auto-generating small parts of an assignment: 65.1\% of high school students considered it ethical, followed by 58.6\% of college students and 43\% of university students (high school vs. university, \(\chi^2(1) = 9.01\), \(p_{adj} = .008\);
college vs. university, \(\chi^2(1) = 6.83\), \(p_{adj} = .027\)).

\subsection{Attitudes}
\insight{\textbf{Key insights:}
\begin{itemize}[leftmargin=*,labelsep=5pt]
    \item Across all groups, students express excitement, optimism, and gratitude toward GenAI, with only a minority feeling fear. 
    \item Students in higher education often report worry and skepticism, while high school students report greater trust and fewer negative feelings.
    \item Higher GenAI usage correlates with more positive and fewer negative attitudes. 
    \item High school students show more stable and increasingly positive attitudes toward GenAI, whereas higher education students report more frequent and more critical shifts in perception.
\end{itemize}
}

\begin{figure}[htbp]
    \centering
    \includegraphics[width=0.47\textwidth]{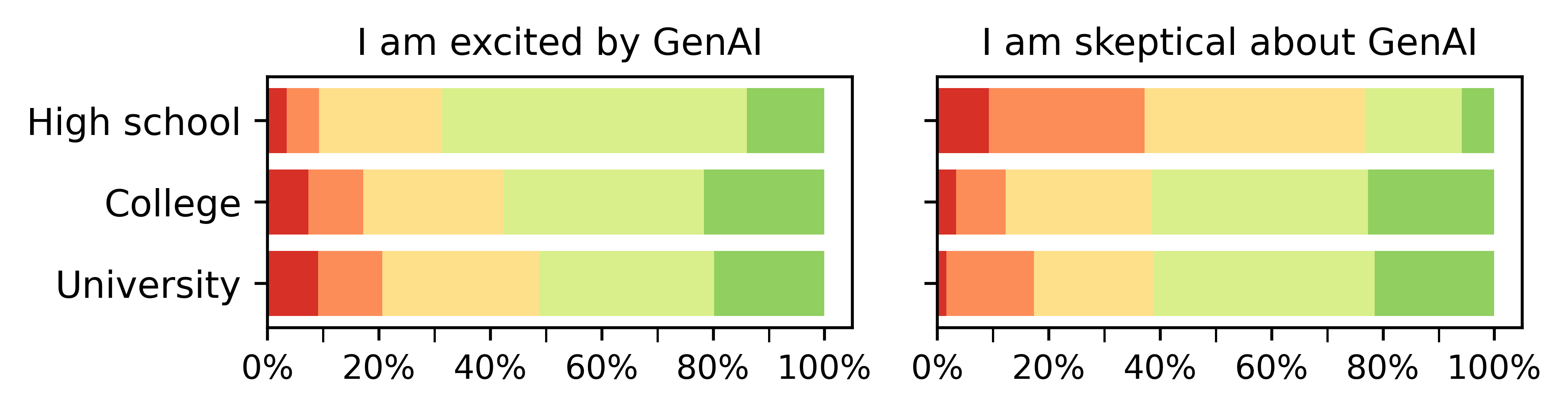}\\[0.5em]
    \includegraphics[width=0.47\textwidth]{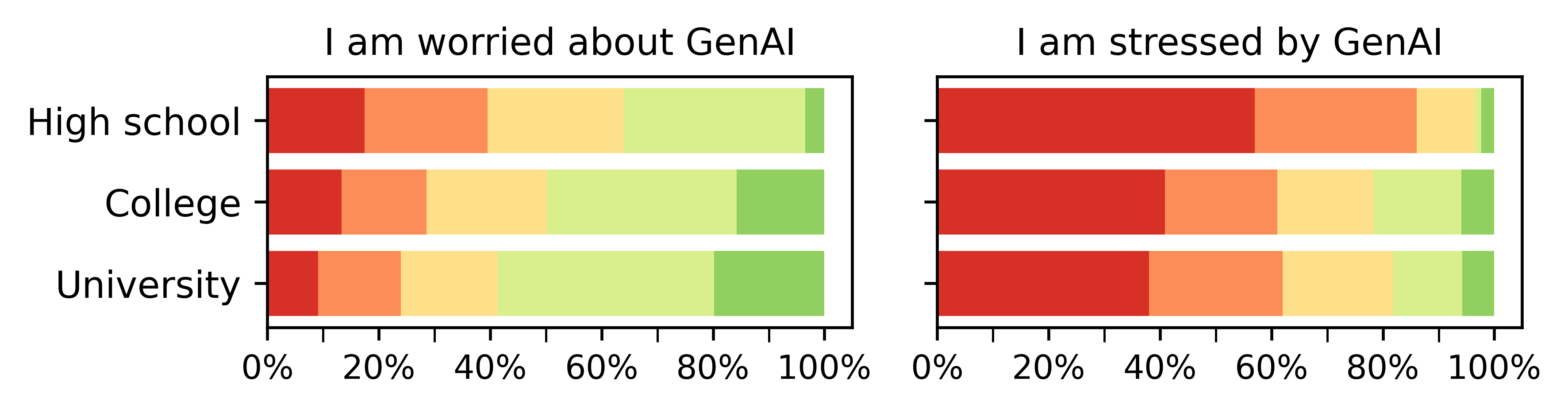}\\[0.5em]
    \includegraphics[width=0.32\textwidth]{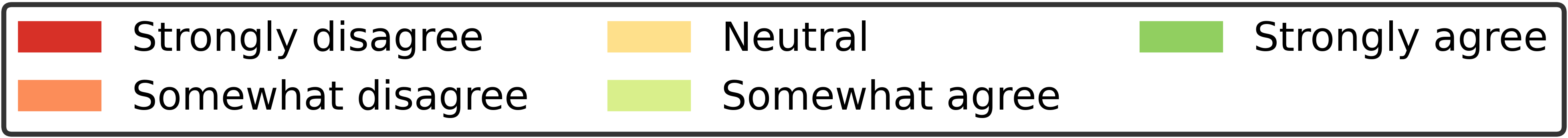}
    \caption{Student attitudes toward GenAI.}
    \label{fig:attitudes}
\end{figure}

Regarding attitudes toward GenAI, we again observe differences between the educational groups. High school students reported greater \textit{trust} in GenAI outputs, with 40.7\% somewhat or strongly agreeing, compared to only 16.8\% of college students and 19\% of university students. Mann--Whitney U tests confirmed these differences (high school vs. college, \(U = 11826\), \(p_{adj} = <.001\); high school vs. university, \(U = 7335.5\), \(p_{adj} = <.001\)). 

To gain a closer look at their attitudes, we asked students to rate their agreement with seven statements of the form “I am \_\_\_\_ by/about GenAI,” reflecting different emotions: \textit{optimistic}, \textit{skeptical}, \textit{worried}, \textit{excited}, \textit{stressed}, \textit{grateful}, and \textit{frightened}. Figure~\ref{fig:attitudes} presents four representative responses. A majority of students across all three educational levels reported feeling \textit{excited} about GenAI: 68.7\% of high school students, 57.7\% of college students, and 51.2\% of university students somewhat or strongly agreed with the statement. Although high school students expressed more \textit{excitement} overall, a Kruskal--Wallis H test showed no statistically significant differences in distributions among the groups.
For the other three attitudes, a clear and significant trend emerges: high school students reported fewer negative feelings toward GenAI compared to college and university students. For example, high school students are the only group where disagreement (39.5\% somewhat or strongly) outweighs agreement (36.1\%) with the statement about feeling \textit{worried} about GenAI. In contrast, more college students agree (49.8\%) than disagree (28.6\%), and the same holds for university students (58.6\% agree vs. 24\% disagree). Pairwise Mann–Whitney U Tests confirmed the stronger attitudes by high school students (high school vs. college, \(U = 7073\), \(p_{adj} = .026\); high school vs. university, \(U = 3677\), \(p_{adj} = <.001\)). A similarly strong and significant trend is observed for the attitudes of feeling \textit{skeptical} and \textit{stressed}.

For the other attitudes, students across all educational groups reported relatively similar levels of emotional responses. More than half of all students in all groups felt \textit{optimistic} about GenAI: 59.3\% of high school students, 61.6\% of college students, and 57.1\% of university students somewhat or strongly agreed. Similarly, students felt \textit{grateful} for GenAI: 55.8\% in high school, 55.1\% in college, and 47.9\% in university. A Kruskal--Wallis H Test showed no difference between the groups. Meanwhile, feelings of \textit{fright} were reported at much lower levels: only 11.6\% of high school students, 20.7\% of college students, and 19\% of university students agreed.

\begin{figure}
    \centering
    \includegraphics[width=1\linewidth]{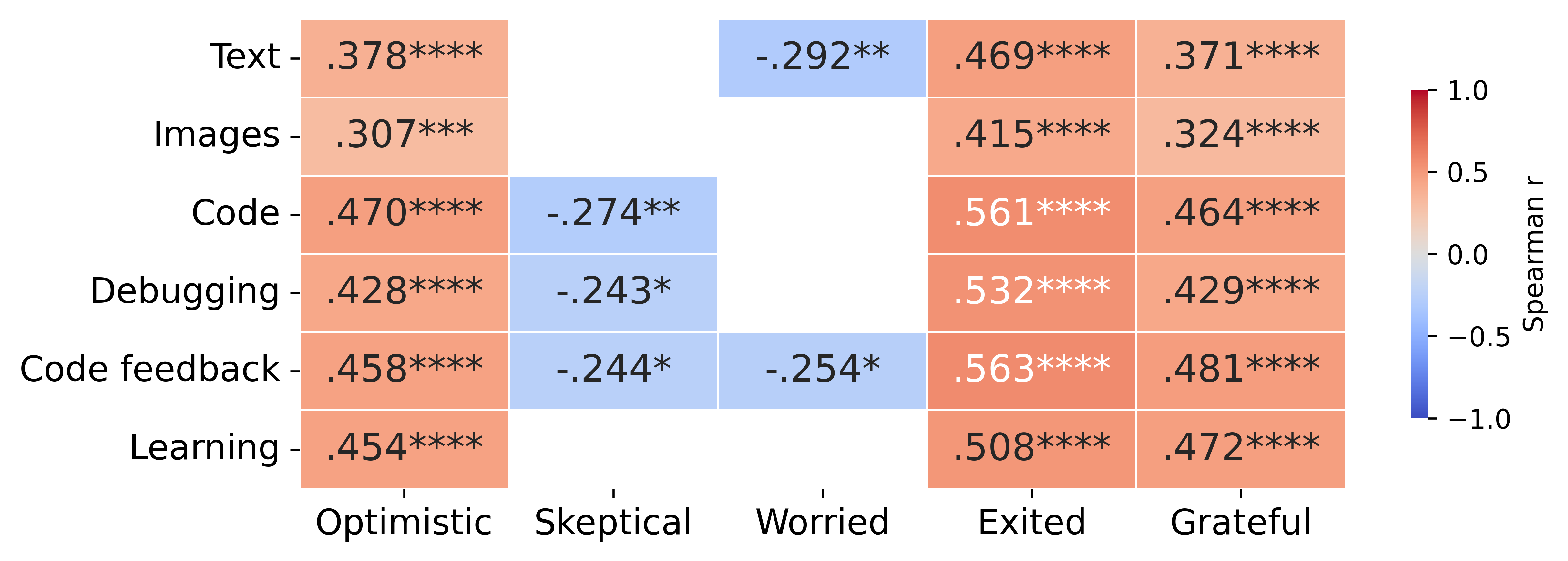}
    \caption{Significant Spearman correlations between GenAI use and attitudes for college students. \\
        \textit{Note.} $^*$ $p < .05$, $^{**}$ $p < .01$, $^{***}$ $p < .001$, $^{****}$ $p < .0001$
        }
    \label{fig:usage_attitudes_correlations}
\end{figure}

Building on earlier findings, we investigated the relationship between GenAI usage and students’ attitudes. Spearman correlation analysis revealed a moderate positive association between the use of GenAI when working with text and an \textit{excited} attitude among high school students, \(r_s = .449\), \(p = <.001\). For college and university students, higher GenAI usage is generally associated with more positive attitudes toward GenAI. Figure~\ref{fig:usage_attitudes_correlations} shows the correlation values for college students. We observe a weak negative association between the use of GenAI for programming tasks and the attitudes of being \textit{skeptical} and \textit{worried}, suggesting that higher usage is linked to a lower likelihood of feeling \textit{skeptical} or \textit{concerned} about GenAI. Additionally, we observe a moderate to strong positive association between general GenAI usage and attitudes of \textit{optimism, excitement, and gratitude}. This suggests that more frequent use is linked to more positive perceptions of GenAI. The correlation map for university students closely resembles that of college students. It shows a moderate negative association between general GenAI usage and attitudes of \textit{skepticism} and \textit{worry}, and a moderate to strong positive association with feelings of \textit{optimism}, \textit{excitement}, and \textit{gratitude}.

Differences in attitudes between students of different educational types also become apparent when examining the open-ended questions. Students were asked to indicate whether their attitude towards GenAI has changed over time and why. In total, 318 students provided a clear answer to this question. In their coded responses, we noticed a difference between high school students and college/university students. For high school students, the majority indicated their attitude towards GenAI had not changed, while the minority indicated it had. Among the college and university students, however, this result was more balanced, with about half of the respondents saying their attitude had changed, while the other half said it did not. From the students in all educational types that reported a change in their attitudes, some students indicated they became more positive about the technology, while others noticed they became more negative. A high school student stating they became more positive indicated, for example, \textit{``They've changed. At first, I was very skeptical and didn't use it because I didn't trust the generated data to be reliable. Now I see much more of its usefulness and use it primarily to explain things to myself or search efficiently."} The same sentiment can be found with students from other types of education. A university student said \textit{``At first I thought GenAI was useless and bad, but because it is constantly being updated and improved, it is becoming more and more useful and helpful."} The results do point towards a difference in the direction of attitude changes between high school and higher education students. Of the limited number of high school students who say their views have changed, most indicated they have become more positive about the technology over time. However, of the college and university students who mentioned a change in their attitude, this seems to be reversed, with more students saying they became more negative.
A university student remarked: \textit{``Generative AI is developing rapidly and is constantly improving, which makes me feel like I'm studying in a risky field. Computer science isn't completely irrelevant, but we don't know what the future holds. This is a pretty scary thought. I'm also worried that critical thinking skills could deteriorate significantly.''}

\subsection{Within courses}
\insight{\textbf{Key insights:}
\begin{itemize}[leftmargin=*,labelsep=5pt]
    \item High school students report more positive teacher attitudes toward GenAI than those in higher education, where negative teacher perceptions are more prevalent, especially in universities.
    \item Most students expressed that GenAI enhances, rather than hinders, their courses.
    \item High school students resort more to friends than students in higher education.
    \item The ranking of GenAI as a help resource for university students in Spring 2025 is similar to the year before.
\end{itemize}
}

In the second part of our survey, we asked high school students to answer with their Informatics classes in mind. For the students in higher education, we asked them to provide the name of one course that involved programming and answer the questions for this course. In 152 cases (55\%) the learning goal of this course is programming, and for 120 responses (44\%) programming is necessary, but not the core learning goal. We excluded 3 responses (1\%) from group analysis since programming was optional in their course. 

\begin{figure}[!bp]
    \centering
    \includegraphics[width=1\linewidth]{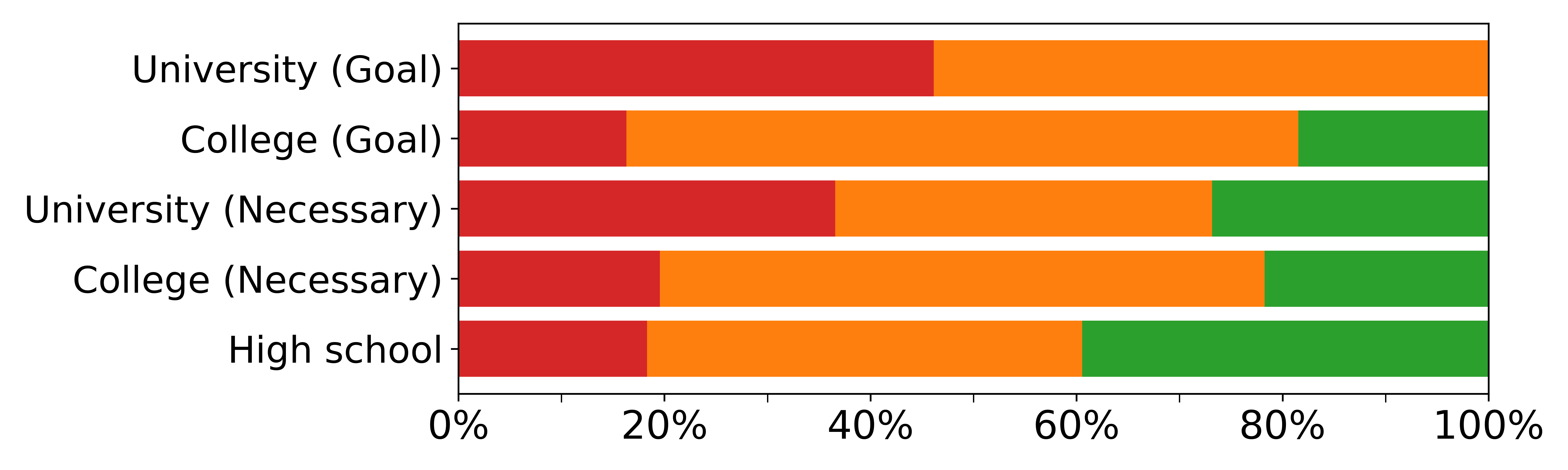}\\[0.5em]
    \includegraphics[width=0.31\textwidth]{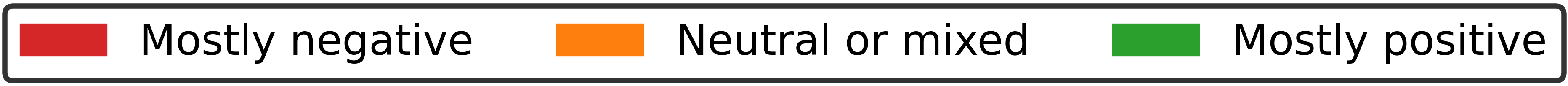}
    \caption{Student-reported teachers' attitudes toward GenAI.}
    \label{fig:teacher_att}
\end{figure}

\subsubsection{Teacher's attitude}
We asked the students what their teacher's attitude was towards GenAI. In universities, 36.8\% of students did not know this, which is at least twice as high as in colleges (17.2\% did not know) and in high school (12.3\% did not know). In general, high school teachers are perceived as more positive (34.6\%), followed by college (16\%), and lastly university (10.4\%). In high school, only 16\% of students think their teacher has a mostly negative attitude towards GenAI. A Chi-Square test confirmed different distributions among the three institutions (\(\chi^2(6) = 45.91\), \(p_{adj} = <.001\)).

Figure \ref{fig:teacher_att} shows the perceived attitude without the unknowns, split between institution and course type. When zooming in, the difference between college and university is influenced by the university courses in which learning programming is the core goal. In these courses, none of their teachers are perceived as mostly positive towards GenAI, and a higher proportion are seen as mostly negative, 46.2\% in universities compared to just 16.3\% in colleges. For college core programming courses, the spread is even, with the majority mixed or neutral (65.2\%) and similar numbers of proponents and opponents. The numbers are similar for courses where programming is necessary but not the core learning goal. In such courses at universities, students feel that their teachers are leaning a bit more towards the positive side, with 26.8\% being mostly positive. Pairwise Mann–Whitney U tests confirmed a significant difference between university and college students in core programming courses (\(U = 1672\), \(p_{adj} = <.001\)), while showing similar distributions for courses where programming is necessary but not the core goal (\(U = 1030\), \(p_{adj} = .426\)).

\subsubsection{Reasons for using or avoiding GenAI}

\begin{figure}[!bp]
    \centering
    \includegraphics[width=0.47\textwidth]{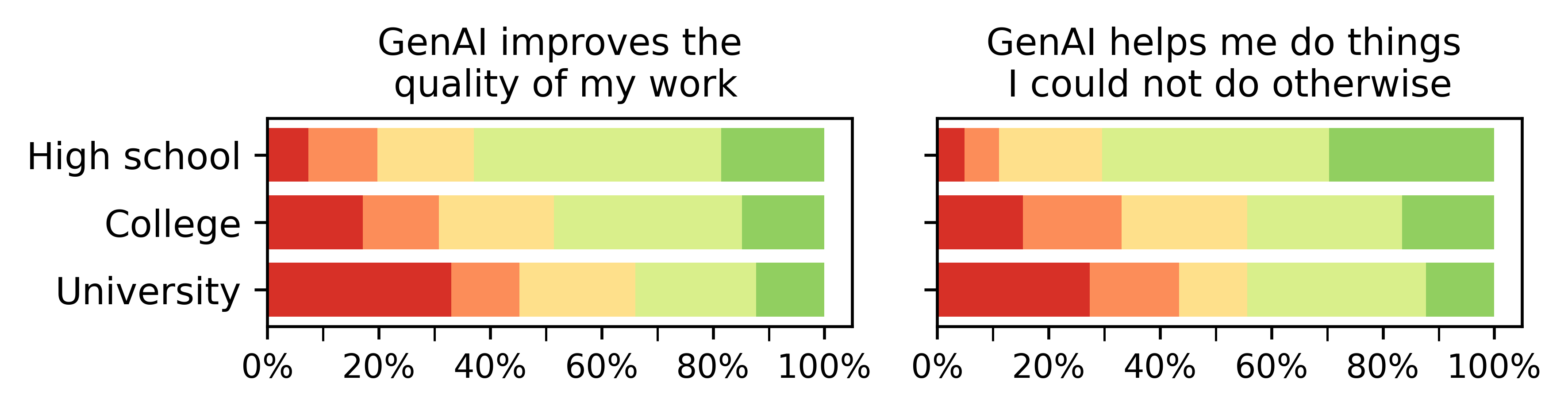}\\[0.5em]
    \includegraphics[width=0.47\textwidth]{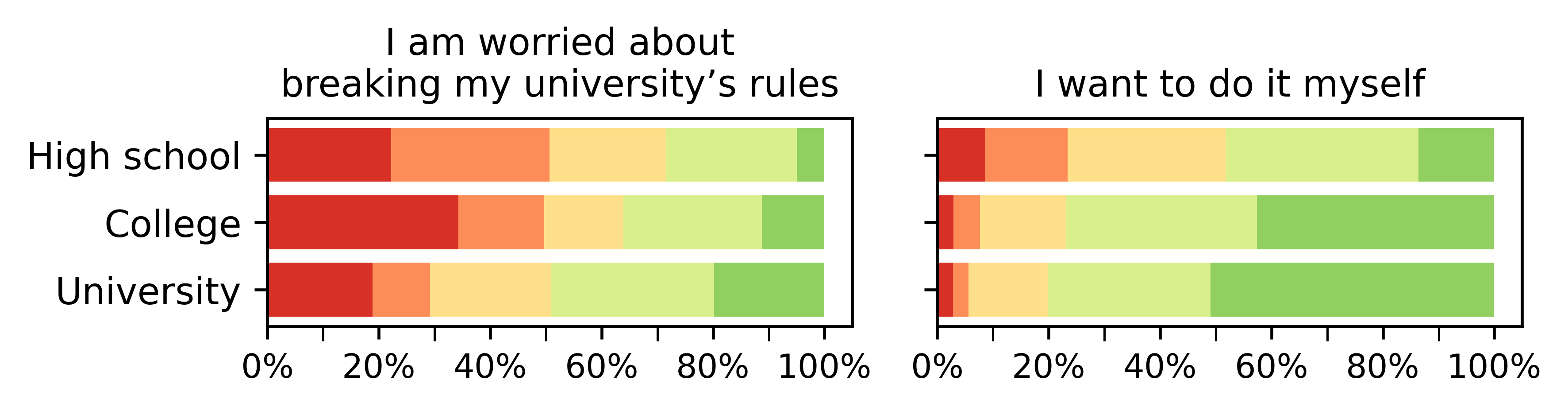}\\[0.5em]
    \includegraphics[width=0.32\textwidth]{Qlikert_legend.png}
    \caption{Reasons for using or not using GenAI.}
    \label{fig:reasons}
\end{figure}

We presented the students with reasons for choosing to use, or not use, GenAI in their course. Multiple Kruskal--Wallis H tests, followed by pairwise Mann--Whitney U tests, confirmed significant differences between the groups in the majority of their reported reasons; four representative examples are shown in Figure~\ref{fig:reasons}.
For the majority of all groups, it makes things easier or faster. Many high school students also answer that it improves the quality of their work (62.9\% somewhat or strongly agree). For college students, this is 48.5\%, and university students only 34\% (\(H(2) = 18.13\), \(p = <.001\), university differing from others).
More than two-thirds (70.3\% somewhat or strongly agree) of the high school students believe GenAI helps them to do things they could not do otherwise, but fewer college and university students claim this--44.4\% in both groups (\(H(2) = 24.45\), \(p = <.001\), high school differing from others).
More higher education students believe that GenAI does not improve their grades, while high school students are more likely to take a neutral stance.
GenAI helping students to feel confident is one of the reasons that high school students agree the least with 19.3\%. For college and university students, this percentage is similar with 30.7\% and 29.2\%. 

Turning to reasons students do \textit{not} want to use GenAI in the course, university students are more worried about breaking their institution's rules: 49\% somewhat or strongly agree compared to 36\% in college and 28.4\% in high school (\(H(2) = 12.52\), \(p = .002\), university differing from others).
For all groups, less than 20\% do not worry about GenAI being inaccurate or making things up.
Only a small percentage (at most 11.3\%) in all groups give not knowing how to use it confidently as a reason.  Higher education students, 80.1\% in university and 76.9\% in college, more often reported wanting to do it themselves, compared to only 48\% of high school students who gave this reason (\(H(2) = 39.69\), \(p = <.001\), university differing from others). Privacy and ethical concerns play an increasing role from high school (21\%), to college (36.7\%), to university (47.2\%).

College students are the largest group to somewhat or strongly agree that GenAI tools played a large role in their learning process during the course (47.9\%), and have accelerated their learning (55.1\%). However, the difference is not significant among the groups.
In an open question asking how students feel using GenAI affected their learning process in the course, these sentiments are echoed as well. The majority of the 242 students who expressed their opinions on this indicated they feel the use of these tools only enhanced their education. Students mentioned the tools help speed up the learning process. They find the tools useful for quickly fixing mistakes, use them to get summaries of course content and ask additional explanations when the teacher is unavailable or their explanations are unclear. A college student described how the tools enhance their learning process: \textit{``When I'm reading something in the book, I ask ChatGPT to provide specific examples that I can play around with to better understand the material. CoPilot speeds up my code typing and prevents me from staring at the screen for too long without doing anything. It accelerates my ability to write code faster without lingering too long.}''
Fewer students indicate they feel the tools only hindered their learning process. The most often provided reason for this being that it makes them learn less. A university respondent, for example, stated: \textit{``I think GenAI hindered my learning process because it required me to think less for myself. If I hadn't used AI, I would have had to think more about my code, and I probably would have remembered it better.''} 
Some the respondents indicate they have experienced both the positive and negative consequences of using GenAI. A high school student states: \textit{``On the one hand, it has helped me understand things faster and better, but on the other hand, I have sometimes used it too easily to solve difficult problems.''}

\subsubsection{Help-seeking}

\begin{figure}[t]
    \centering
    \includegraphics[width=0.8\linewidth]{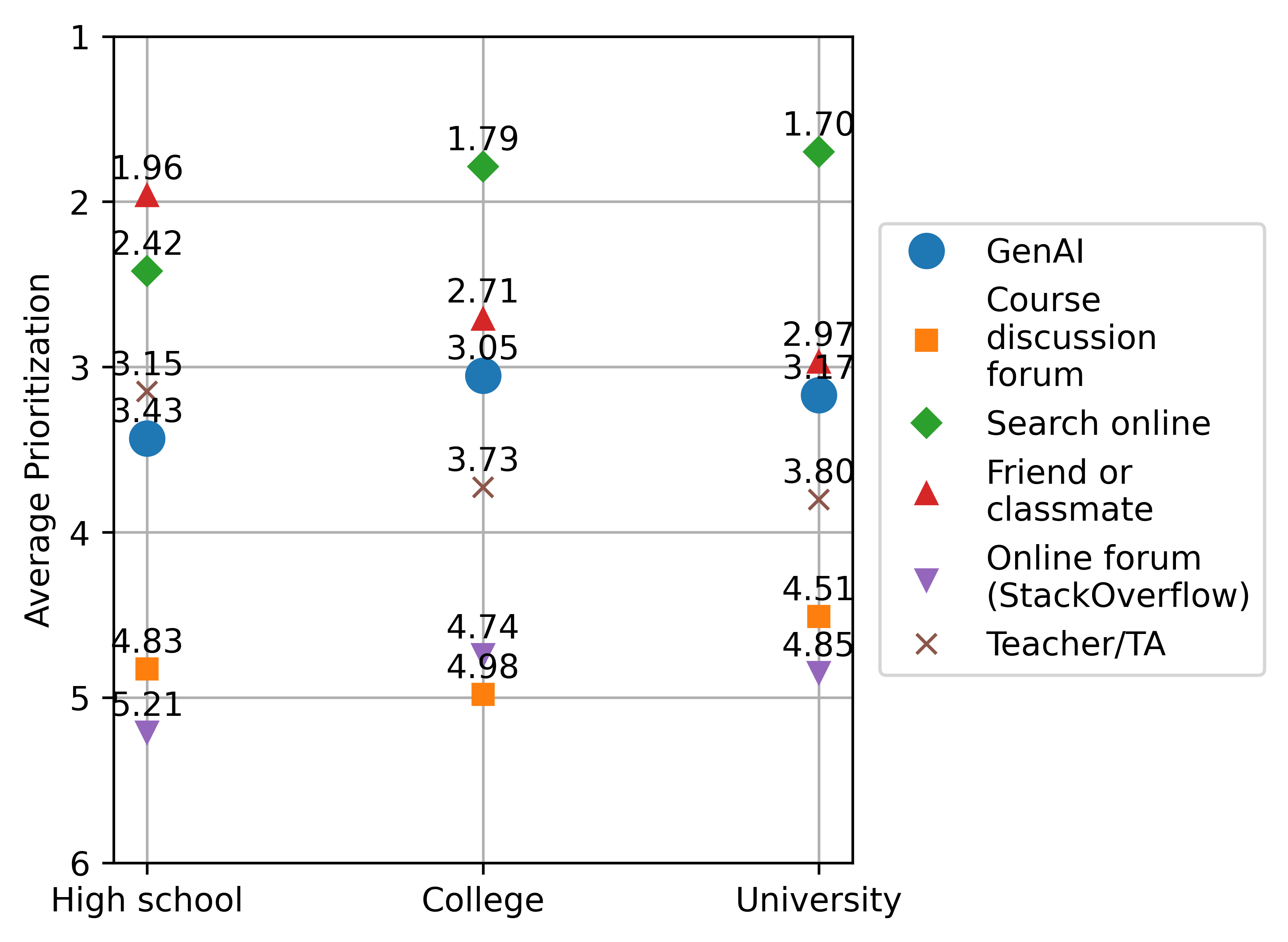}
    \caption{Average student rankings of help resources.}
    \label{fig:help}
\end{figure}

Finally, we asked how students looked for help within the course. Figure \ref{fig:help} shows the order in which the different groups ranked different help resources.
Notably, high school students ranked friends and classmates the highest. Apart from that, online search was still ranked the highest. GenAI was ranked behind online search and friends, and for high school students, their teacher plays a more important role.

In the 2023--24 academic year the same question was asked three times throughout the year in another survey conducted among European university students \cite{keuning2024students}. 
If we compare their last measurements for university students to our university students, online search, GenAI, and friends are ranked slightly lower on average, and  the other resources slightly higher, but the differences are minimal. The rise of GenAI as a help resource that was noticed in 2023--24 seems to have stabilized.

\subsection{Influence on study choice}

\insight{\textbf{Key insight:} The rise of GenAI has a minimal influence on high school students' choice to study computing.
}

We asked the high school students whether they have considered studying Computer Science or a related study, and if the rise of GenAI changed this decision. In total 22.5\% still considers it, 27.5\% does not anymore, 11.2\% does not know, and 38.8\% does not consider it.
There is no major influence of GenAI on this decision, similar but small numbers of 80 students that answered this question reported that it has influenced their decision in a positive (11\%) or negative (10\%) way. For most students, there was no influence (79\%).

\section{Discussion}
\label{discussion}

In this section, we provide an answer to our research question by discussing the main findings that describe how each group of students uses and perceives GenAI. Additionally, we relate our findings to previous surveys and other studies. Finally, we discuss the implications and limitations of our study.

\subsubsection*{Increased usage} In early 2023, only 15.8\% of university students used code generators frequently \cite{smith2024early}. In contrast, our study found that 49.6\% of university and 67.1\% of college students now generate code at least weekly--likely due to the integration of coding features into chatbots and improvements in their coding capabilities \cite{prather2023robots}. Despite frequent use of GenAI for certain tasks, students remain skeptical about GenAI replacing teachers. In the survey by \citet{prather2023robots}, 42\% of students believed GenAI could guide as effectively as a teacher. In our survey, fewer college (32.5\%) and university (37.2\%) students agreed. However, open-ended responses suggest that students view GenAI as a useful supplement for explanations, indicating a more complementary role for GenAI. Comparing the GenAI usage among high schools with the findings of \citet{LEE2024100253}, learning remains one of the primary tasks, reported 57\% in our survey, compared to 39.29\% to 42.65\% in theirs. Looking at the ethical usage of GenAI, most students in the same study \cite{LEE2024100253} agreed that it is unethical to submit assignments completely generated by GenAI. In our study, we asked the same question but added the condition that you do understand what you are submitting. Here, although most students still agree this is unethical, about 35\% of students say they would find this ethical. Finally, high school students admit to having misused or having seen misuse of GenAI more frequently than our college and university respondents.

\subsubsection*{To use or not to use...} Our participants gave several reasons to use GenAI, or not, in the computing course they had in mind. We can compare these results to a 2024 large-scale survey on student perspectives on AI in higher education by \citet{Chung2024}, from which we adopted this question. This study was conducted at four Australian universities in the second semester of 2024, with responses from over 8,000 students. In this previous study, 75\% of students indicated using GenAI because they think it improves the quality of their work. Note that the results report on agree/disagree about the reasons, which we assume are the students who somewhat and strongly agree/disagree.\footnote{\url{https://aiinhe.org/wp-content/uploads/2024/10/aiinhe_surveyinsights.pdf}} This percentage is much higher than 34\% of our university computing students and 48.5\% of college students. This might be due to the nature of computing students' work, which is often focused on code, proofs, etc., that may require more skills to manage effectively, as opposed to generating other artifacts (e.g., reports). Also, a little less than half of our university and college students think GenAI can help them do things they could not do otherwise, versus 70\% of the students in the previous study.

\subsubsection*{Feelings about AI}
We can also compare these results to those of \citet{Chung2024}.
The percentage of students feeling excited about GenAI in that study is roughly equal to our university group. However, our university computing students report being worried more often (59\% compared to 47\%), but the percentage of students feeling stressed is lower (18\% versus 27\%).
Students are similarly skeptical (61\% versus 56\%), optimistic, and grateful. The Australian students, however, were much more frightened of GenAI (28\% versus 19\% in our case). It seems the more extreme emotions were felt less among our university students, but it would require a more in-depth analysis to uncover the reasons for this.

\subsubsection*{Resorting to friends or AI for help?}
High school students consult their friends the most when they need help. This number one position is replaced by online search for students in higher education; however, it is hopeful to see that friends and classmates still play an important role here as well. \citet{hou2025roads}'s interview study with American students revealed a more pessimistic view, in which students help each other less and just ``chatgpt'' all of their questions, leading to problems with isolation and motivation. We did not see clear evidence for this ``erosion'' of social learning, but it would be interesting to replicate these in-depth interviews in a European setting. Perhaps a more competitive climate, or more exposure to GenAI tools, could be the cause for differences.

\subsubsection*{The computing students of the future}

We did not find a major effect of the emergence of GenAI on the study choice of our high school students. We should note that the students we asked were already taking the elective informatics program, so they are likely interested in the topic. Their teachers also appear to have a more positive or neutral attitude towards GenAI, and the media in the Netherlands pays attention to both the possible benefits of AI, as well as the problems and potential harm.

\subsubsection*{Implications} We observe that students from the different educational settings show distinct patterns in how they use and perceive GenAI. However, the reasons for these differences remain unclear. One possible explanation is the specific characteristics (e.g., structural) of each context. For example, high school students use GenAI less for programming-related tasks, which aligns with the fact that they have one informatics course in their study program. College students report using GenAI more frequently for programming-related tasks, along with more permissive institutional policies and more positive teacher attitudes, compared to university students. This may reflect the more practical nature of their studies. For high school students, it is also unclear whether their more permissive and less negative attitudes stem from their younger age and still-developing perspectives, from a limited understanding of AI and its implications, or from being part of a generation that is not only digitally native \cite{bennett2008digital} but also growing up as “AI natives.”
These positive and permissible attitudes of high school students can have implications for educators across all educational types. For high school educators, it may be advisable to create interventions that build high school students' GenAI literacy, so that students understand how these tools work and their limitations. For college and university educators, it is important to know that students coming from high school place much trust in GenAI tools, and what experiences they have using them for tasks like text generation. Beyond supporting students' knowledge and skills in GenAI use, it is essential to clearly communicate the expectations and boundaries regarding the use of GenAI to students transitioning from high school to higher education.

\subsubsection*{Limitations}
The students self-reported on their usage, views, and consequences of GenAI use. These responses could potentially not represent reality, due to various reasons: they might not have seriously answered and were only interested in winning a gift voucher, they might not realistically assess their habits and consequences of using GenAI, and lastly, even though the survey was anonymous, they might have given untruthful answers to some of the more delicate questions. 
The respondents could also be more than average interested in GenAI, leading to stronger positive and negative views. Additionally, the third open-ended question was coded by a single author, which could have led to reduced reliability and potential bias in the findings. Finally, our study was conducted in a single European country. Findings might generalize to similar contexts, but we expect differences in other continents and countries.

\section{Conclusion and Future Work}
\label{conclusion}
In this study, we have explored how 410 students from high school, college, and university report using and perceiving GenAI in the context of computing education. Some key insights include: (1) high school, college, and university students use GenAI in different ways; (2) high school and college students hold more permissive views on the ethics of using GenAI; (3) higher GenAI usage is linked to more positive and fewer negative attitudes; and (4) GenAI has little influence on high school students’ decision to study computing.

As future work, we plan to administer a follow-up survey to the students who participated in the initial study, allowing us to track changes in their GenAI usage and perceptions over time. Additionally, we share our instruments to encourage other researchers to conduct our study in other countries.
\balance
\bibliographystyle{ACM-Reference-Format}
\bibliography{refs}

\end{document}